%% file: main.tex
\documentclass[twocolumn,preprintnumbers,superscriptaddress,amsmath,amssymb,prb]{revtex4}

\usepackage{graphicx}
\usepackage{lipsum}
\usepackage{dcolumn}
\usepackage{subfigure}
\usepackage{bm}
\usepackage[utf8]{inputenc}  
\usepackage{comment}

\usepackage[cmyk,dvipsnames]{xcolor}
\definecolor{pacificb}{HTML}{1CA9C9}
\newcommand{\AnnaD}[1]{\textsf{\textcolor{teal}{[\textit{#1}]}}}
\newcommand{\BS}[1]{\textsf{\textcolor{red}{[\textit{#1}]}}}

\begin{document}

\title{Spin transport properties in a topological insulator sandwiched between two-dimensional magnetic layers}

\newcommand{\KTH}{Department of Applied Physics, School of Engineering Sciences, KTH Royal Institute of Technology, 
AlbaNova University Center, SE-10691 Stockholm, Sweden}
\newcommand{\SeRC}{Swedish e-Science Research Center (SeRC), KTH Royal Institute of Technology, SE-10044 Stockholm, Sweden}
\newcommand{\Uppsala}{Department of Physics and Astronomy, Uppsala University, Box 516, SE-75120 Uppsala, Sweden}
\newcommand{\WISEKTH}{Wallenberg Initiative Materials Science for Sustainability (WISE), KTH Royal Institute of Technology, SE-10044 Stockholm, Sweden}

\author{N. Pournaghavi}   \email{nezhat@kth.se}
    \affiliation{\KTH}

\author{B. Sadhukhan}  
    \affiliation{\KTH}

    \author{A. Delin}  
    \affiliation{\KTH}
    \affiliation{\SeRC}
    \affiliation{\WISEKTH}

\date{\today}

\begin{abstract}
Nontrivial band topology along with magnetism leads to different novel quantum phases. When time-reversal-symmetry is broken in three-dimensional topological insulators (TIs) by applying high enough magnetic field or proximity effect, different phases such as quantum Hall or quantum anomalous Hall(QAH) emerge and display interesting transport properties for spintronic applications. The QAH phase displays sidewall chiral edge states  which leads to the QAH effect. In a finite slab, contribution of the surface states depends on both the cross-section and thickness of the system. Having a small cross-section and a thin thickness leads to direct coupling of the surfaces, on the other hand, a thicker slab results in a higher contribution of the non-trivial sidewall states which connect top and bottom surfaces. In this regard, we have considered a heterostructure consisting of a TI, namely Bi$_2$Se$_3$ , which is sandwiched between two-dimensional magnetic monolayers of CrI$_3$ to study its topological and transport properties. Combining DFT and tight-binding calculations along with non-equilibrium Green’s function formalism, we show that a well-defined exchange gap appears in the band structure in which spin polarised edge states flow. We also study the width and finite-size effect on the transmission and topological properties of this magnetised TI nanoribbon.  
\end{abstract}

\maketitle

\input{01_Introduction}
\input{02_Method}
\input{03_Results}

\input{04_Conclusions}
\bibliography{Some_refs.bib}

\end{document}

%% file: 01_Introduction.tex
\section{Introduction}
%

Three dimensional topological insulators (3D TIs) are characterized by an insulating gap in the bulk and gapless 2D surface states at the termination surfaces inside this gap, which are topologically protected by time-reversal symmetry (TRS)\cite{hasan2010, XLQi, Ando2013,Bansil2016}. In the presence of TRS, the topological surface states consist of spin-momentum locked states forming a single massless Dirac cone, whose existence is guaranteed by the bulk ${\cal Z}_2$ topological invariant. 
In this system, at each edge, helical states with opposite spin propagate in the opposite directions. Introducing magnetic order in TIs breaks TRS and results in different topological phases. Breaking TRS can be for instance due to the high enough magnetic field which leads to the quantum Hall effect (QHE), or introducing spontaneous magnetization which leads to the quantum anomalous Hall effect (QAHE). 

When TRS is broken at the surface of a TI, a half-integer quantum anomalous Hall conductivity $\pm e^2/2h$ ($h$ is the Planck constant and $e$ is the electron charge) arises at that surface\cite{Fu2007, Qi2008, Essin2009, Vanderbilt}. In a TI thin film in a Hall bar geometry, the resulting topological state depends on how TRS is broken at the two 
surfaces. When the magnetization at the top and the bottom  surfaces points in the same direction and the chemical potential is inside the exchange gap, the system is in the Chern insulator (CI) state. This phase is characterized 
by a non-zero integer Chern number $\cal C$ which displays the QAHE where due to the emergence of chiral edge states on the film side walls, the Hall conductance is quantized $\sigma_{\rm H} = {\cal C} e^2/h$ \cite{Yu2010,Chang2013,Liu2016,Chang2016}.

One approach to generate magnetism at the surfaces of a TI thin film 
is to exploit the interfacial proximity with an adjacent film of a 
magnetic insulator or semiconductor\cite{Luo2013, Eremeev2013, Wei2013}. This method is favorable over doping the system with magnetic impurities, since the properties of the topological system will be less modified. The crucial issues here arethe selection of the proper magnetic materials, and the nature of their coupling with the TI film.
The magnetic layer should be able to generate a sizable exchange gap; at the same time, the interface hybridization should not be too strong to avoid damaging the Dirac surface states or shifting them 
away from the Fermi level. For this purpose, over the past few years, several magnetic TI heterostructures have been proposed theoretically and realized 
experimentally\cite{Yang2013, Lang2014, Katmis2016, Tang2017, Hirahara2017, Zhu2018, Mogi2019a}. 
The majority of these consist of ferromagnetic (FM) insulators, but a few examples with 
antiferromagnetic (AFM) materials have been considered\cite{Luo2013, Eremeev2013,he_PRL2018, Yang2020,2102.01632,Mogi2019b}. 
Despite all this effort, realization of the CI in these heterostructures is still quite challenging. There are several reasons for this issue however the finite-size effect is one of the main factors that should be taken into account when studying this phase.

In this paper, we employ density functional theory (DFT) along with tight-binding model to study the electronic and topological properties of an FM/TI/FM tri-layer heterostructure, where the 2D ferromagnetic $CrI_3$ is used to magnetize the Dirac surface states of Bi$_2$Se$_3$ \cite{Huang2017}. $CrI_3$ is a 2D van der Waals ferromagnetic monolayer with high Curie temprature (45 K)\cite{Huang2017}, thus,it is a very promising candidate to magnetize surface states of Bi$_2$Se$_3$ without destroying its topological character.
The goal of this work is to investigate the physical conditions under which the emergence of the edge states in the Chern insulator phase can be detected.


%% file: 02_Method.tex
\section{Method}\label{method}
 \subsection{Density functional calculations}
To study the electronic and topological properties of the CrI$_3$-Bi$_2$Se$_3$-CrI$_3$ heterostructure, we have constructed a periodic slab consisting of six quintuple layers (QLs) of Bi$_2$Se$_3$ slab which is sandwiched between inversion-symmetric CrI$_3$ monolayers. To match the lattice constants one needs to consider $a\sqrt{3}\times\sqrt{3}$ supercell of Bi$_2$Se$_3$ in the lateral plane to match the CrI$_3$ monolayer \cite{Hou2019}. We have included a 15~Å vacuum between adjacent slabs to avoid any couplings between them. Placing CrI$_3$ on top of the TI  can be done in several way, but the configuration in which the Cr ions are placed on top of the Se atoms is energetically favorable\cite{Hou2019}. 
Therefore, all  the self-consistent calculations in this work are performed using this setup. All the DFT calculations are performed by employing the Vienna {\it ab initio}
Simulation Package (VASP)\cite{vasp1,vasp2}, and using the Perdew-Burke-Ernzerhof
generalized gradient approximation (PBE-GGA) for the exchange correlation
functional\cite{Perdew1996}. We have first relaxed the crystal structure for
both the cell parameters and the atomic positions using a $k$-mesh of size 6
$\times$ 6 $\times$ 1 until the stress on the cell and the average forces on the atoms
are 0.02 eV/\AA. The final relaxed structure is then used to study the electronic properties with the 
inclusion of the spin-orbit coupling (SOC); for this part we use a larger $k$-mesh of 12
$\times$ 12 $\times$ 1 to improve the accuracy of the calculations. To
incorporate the effect of correlations at the transition metal Cr atoms,  all
self-consistent calculations are performed using GGA+U for Cr atoms with effective on-site exchange interactions of 0.9 eV and effective on-site coulomb interactions of 3 eV.

\subsection{Four band effective and tight-binding models}
For the topological studies of this system, we have to construct a real-space Hamiltonian in the basis of the Wannier states for the low-energy bands. Constructing Wannier states of this complex system from DFT calculations is computationally very challenging, particularly with SOC. Therefore we used two model Hamiltonians to describe the topological properties and transport for the CrI$_3$-Bi$_2$Se$_3$-CrI$_3$ heterostructure.  We used a four-band effective Hamiltonian in momentum space to analysis the topological properties and a tight-binding model to describe accurate transport properties.

\subsubsection{Low-energy effective model}

It is possible to characterize the low-energy long-wavelength properties of the system using a simple effective Hamiltonian, since the topological nature is determined by the physics near the $\Gamma$ point. Therefore, we employ here a low energy effective $k. p$ model to describe the topological properties for the CrI$_3$/Bi$_2$Se$_3$/CrI$_3$ slab heterostructure.  The effective  $k.p$ Hamiltonian is given by \cite{doi:10.1126/sciadv.aaw1874, doi:10.1146/annurev-conmatphys-031115-011417, doi:10.1126/science.1187485}
\begin{eqnarray}\label{model0}
H_{k.p}=&&{\mathcal{H}}_{\mathrm{surf}}+{\mathcal{H}}_{\mathrm{Zeeman}}+{\mathcal{H}}_{\mathrm{Interface}}
   \nonumber
   \\
      &&= \epsilon(k)I_{4\times4}+ \left(\begin{array}{cccc}
   0 & iv_F k_- & m(k) & 0\\
   -iv_Fk_+ & 0 & 0 & m(k)\\
   m(k) & 0 & 0 & -iv_Fk_-\\
   0 & m(k) & iv_Fk_+ & 0
   \end{array}\right)
   \nonumber
   \\
   &&+\left(\begin{array}{cccc}
   \Delta & 0 & 0 & 0\\
   0 & -\Delta & 0 & 0\\
   0 & 0 & \Delta & 0\\
   0 & 0 & 0 & -\Delta
   \end{array}\right)
 +\left(\begin{array}{cccc}
   V_p & 0 & 0 & 0\\
   0 & V_p & 0 & 0\\
   0 & 0 & -V_p & 0\\
   0 & 0 & 0 & -V_p
   \end{array}\right)
   \nonumber
   \\
\end{eqnarray}
where the basis are $|t\uparrow\rangle$, $|t\downarrow\rangle$, $|b\uparrow\rangle$ and $|b\downarrow\rangle$, and $t$, $b$ denote the top and bottom surface states and $\uparrow$, $\downarrow$ represent the spin up and down states, respectively.  Here $\epsilon(k) = A(k_x^2+k_y^2)$,  $v_F$ is the Fermi velocity and  $k_{\pm} = k_x {\pm} ik_y$ are the wave vectors respectively.  $\Delta$ is the exchange field along the $z$ axis introduced by the FM ordering from CrI$_3$ monolayer at the interface.  Here, $\Delta\propto \langle S\rangle$ with $\langle S\rangle$ the mean field expectation value of the local spin.  $m(k)$ describes the tunneling effect between the top and bottom surface states of the Bi$_2$Se$_3$ slab and is given by $k$, $m(k)=M-B(k_x^2+k_y^2)$.  When the slab thickness is very high, the top and bottom surface states are well separated spatially i.e $m(k)\approx0$.  However, with the reduction of the film thickness, $m(k)$ becomes finite and $\left|M\right|<\left|\Delta\right|$ guarantees that the system is in the topological state.  ${\mathcal{H}}_{\mathrm{Interface}}(k_x,k_y)$ describes the inversion asymmetry between the top and bottom surfaces due to the existence of a substrate (i.e., the CrI$_3$ monolayer) where V$_p$ represents the magnitude of the asymmetric interfacial potential that can occur in experiments by the mis-alignment of layers and we keep it as a parameter that can affect topological properties.

\par The parameters of the effective model are obtained from fitting the energy spectrum of the effective Hamiltonian with DFT-calculated bands of CrI$_3$/Bi$_2$Se$_3$/CrI$_3$ slab near the $\Gamma$ point. our fitting leads to $v_f$ = 2.395403 eV.Å; M = 0.001086 eV; B = 2.996947 eV. Å$^2$; $\delta$ = 0.007838 eV; A = 19.000841 eV.Å$^2$.

\subsubsection{Tigh-binding model}

\begin{figure} [ht]
\includegraphics[scale=0.6]{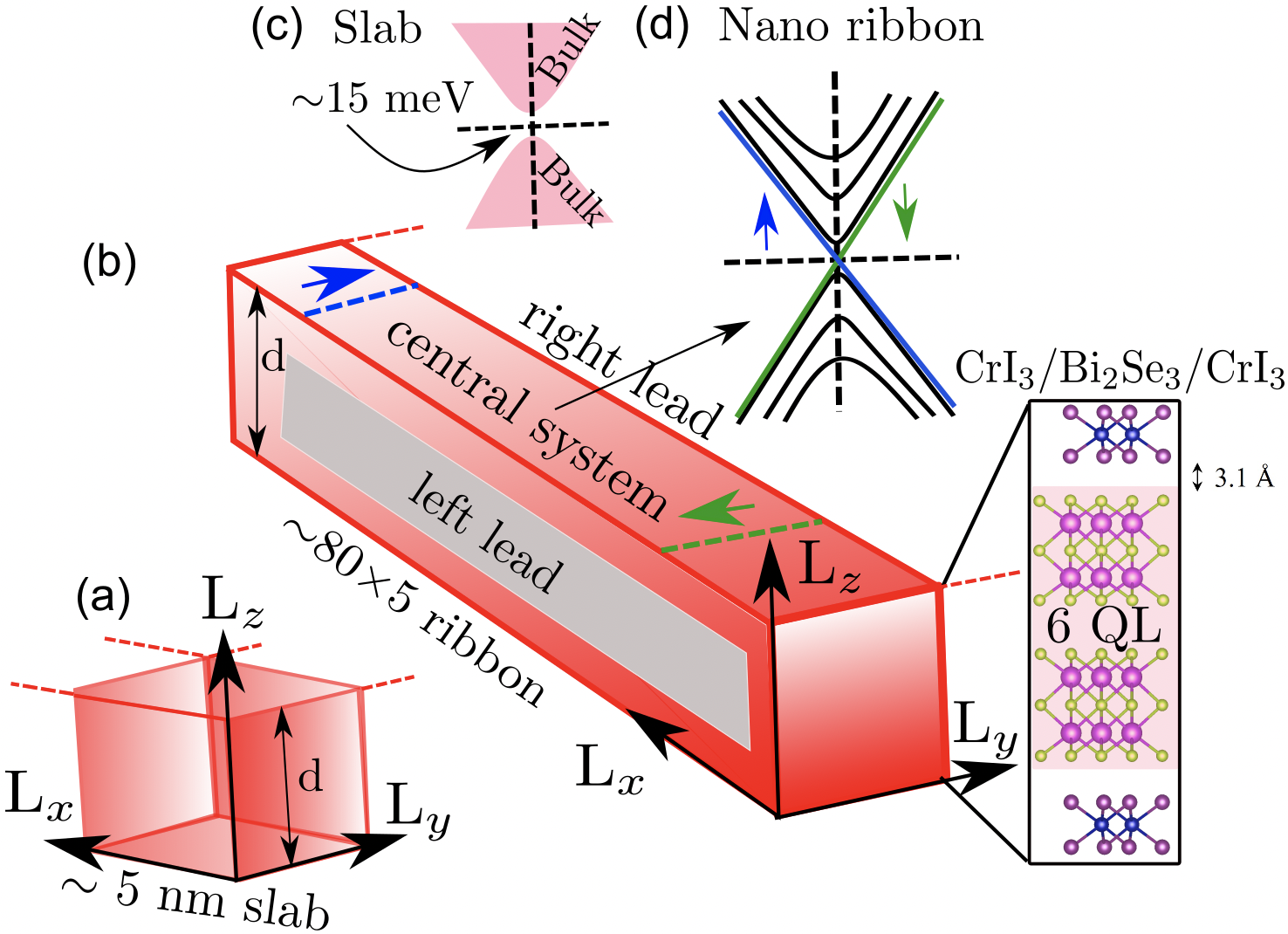}
\caption{Schematic view of the CrI$_3$/Bi$_2$Se$_3$/CrI$_3$ heterostructure in a) the slab geometry where d is the thickness, and b) the nano ribbon geometry. The corresponding band strcutures of the slab and nano ribbon systems in the presence of spin-orbit coupling are plotted in c) and d), respectively.
\label{Model}
}
\end{figure}

Our study further elaborated
a real space tight binding model Hamiltonian mimicking ribbon geometry of  CrI$_3$-Bi$_2$Se$_3$-CrI$_3$ heterostructure to describe the topological properties for spin transport.
In order to have the possibility to study the width dependency of emerging of the edge states we have built an effective TB Hamiltonian containing the effect of the exchange field obtained from DFT calculation, which is then used to consider a quasi 1D nanoribbon geometry with different widths. The acceptable accuracy of the TB Hamiltonian is determined by the criterion that the bands obtained from this Hamiltonian should be a good match with the corresponding bands obtained from the full DFT calculations. The goal is to see the emergence of chiral edge states on the sidewalls of the nanoribbon when the system is in the CI phase. This method is more feasible and efficient than doing several first-principle calculations for different nanoribbons which qualitatively gives the same results.
Using an atomistic tight-binding (TB) model, we have first constructed a pristine
Bi$_2$Se$_3$ slab of 6 QLs fig.(\ref{Model}(a) and zoom-in structure in fig.\ref{Model}(b). To break the TRS, we have added an exchange field at the
top and the bottom surface layers to mimic the effect of the magnetic film on
the surface states. Since the strength of the exchange field determines the size
of the gap at the Dirac point, we have chosen an exchange field that opens up a gap of $\approx 15$ meV, the same order of the gap that we have obtained from the DFT calculations.
To model the Hamiltonian of Bi$_2$Se$_3$ we have used the following $sp^3$
TB model\cite{Kobayashi2011, Pertsova2014, Pertsova2016} 
\begin{equation}\label{eq:2} \begin{aligned} H_C &= \sum
_{ii',\sigma\alpha\alpha'}t_{ii'}^{\alpha\alpha'}e^{i{\bf k}\cdot {\bf
r}_{ii'}}c_{i\alpha}^{\sigma\dag}c_{i'\alpha'}^{\sigma}\\ &\quad + \sum
_{i,\sigma\sigma',\alpha\alpha'} \lambda_i <i,\alpha,\sigma|{\bf L} \cdot {\bf
S}| i,\alpha',\sigma'> c_{i\alpha}^{\sigma\dag}c_{i\alpha'}^{\sigma'}\\ &\quad +
\sum _{i,\sigma ,\alpha} M_i c_{i\alpha}^{\sigma\dag } {\sigma}_{z}^{\sigma
\sigma}c_{i\alpha}^{\sigma}, \end{aligned} \end{equation}

In the first term, $t_{ii'}^{\alpha\alpha'}$ are the Slater$-$Koster parameters
for the hopping energies.  $c_{i\alpha}^{\sigma\dag}(c_{i\alpha}^{\sigma})$ is
the creation (annihilation) operator for an electron with spin $\sigma$ and the
atomic orbital $\alpha \in$ ({$s, p_x, p_y p_z$}) at site $i$. $k$ is the
reciprocal-lattice vector that spans the BZ. $i^\prime \neq i$ runs over all the
neighbors of atom $i$ in the same atomic layer as well as the first and second
nearest-neighbor layers in the adjacent cells ($r_{ii\prime}$ represents the
vector connecting two neighbor atoms).  In the second term, the on-site SOC is
implemented in the intra-atomic matrix elements \cite{Walter1999}, in which
$|i,\alpha,\sigma>$ are spin- and orbital-resolved atomic orbitals. ${\bf L}$
and ${\bf S}$ are the orbital angular momentum and the spin operators,
respectively, and $\lambda_i$ is the SOC strength \cite{Kobayashi2011}. The last
term indicates the exchange field to break the TRS. We assume M=0.05 eV only
for the surface atoms yielding a 15 meV surface gap.

%% file: 03_Results.tex
\section{Results}
\subsection{Construction and  electronic properties of heterostructure}

In the presence of an exchange field in the vicinity of a topological insulator, an exchange gap appear at the TI surface states. The size of the gap depends on the magnitude of the magnetization which is induced as well as the coupling between the adjacent layers in the heterostructure.  Fig.~\ref{Model} shows the configuration that we have considered which consists of a 6QL Bi$_2$Se$_3$ that is sandwiched between two CrI3 monolayers.

\begin{figure} [htb]
\includegraphics[scale=0.4]{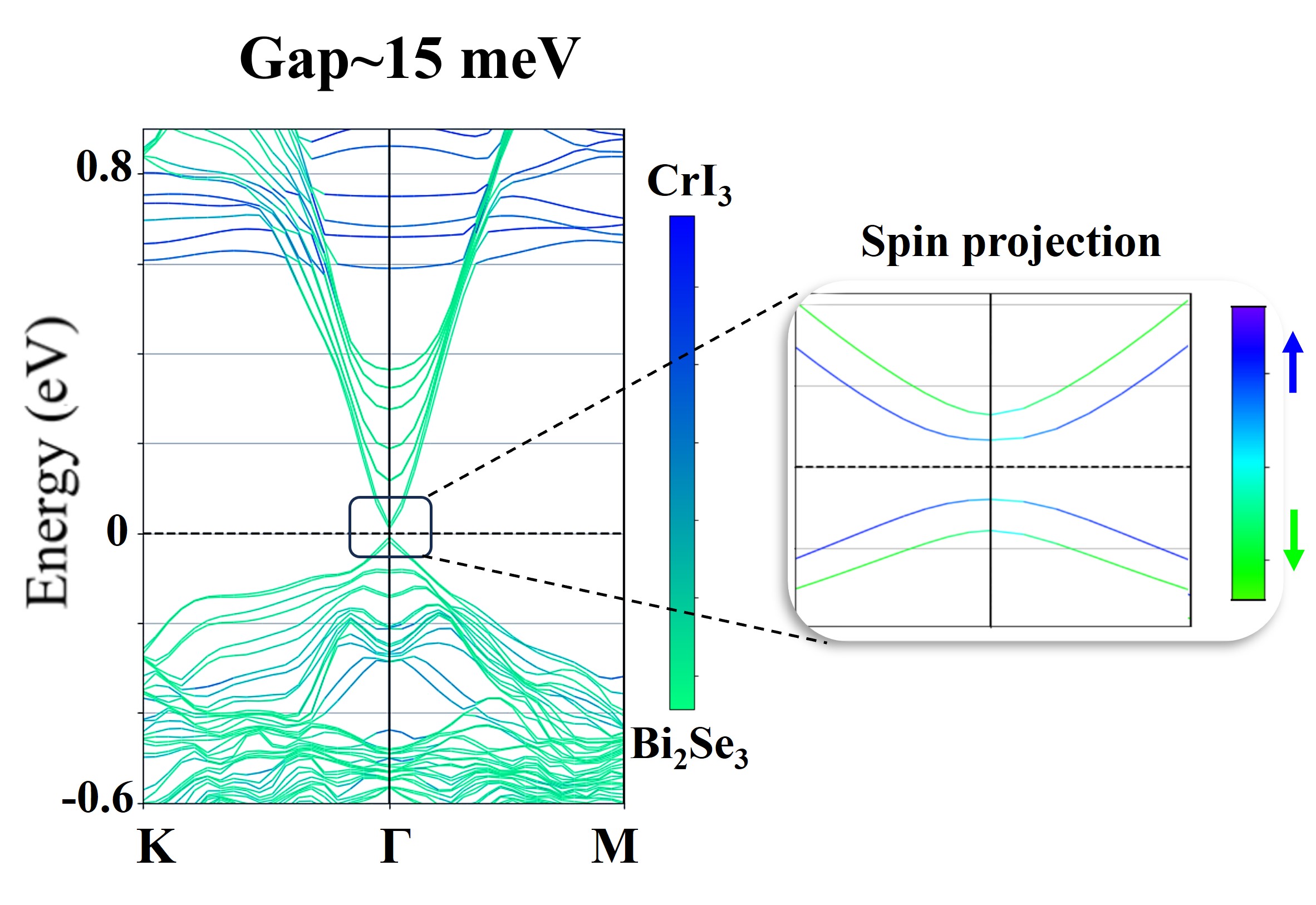}
\caption{Element projected 
band structure in the presence of spin-orbit coupling. The zoom-in figure shows the spin projection of the low energy bands.
\label{DFTband}
}
\end{figure}

As it is shown in fig.~\ref{DFTband}, the induced exchange gap is about 15 meV. In this configuration, the spin character of the conduction surface states close to the Fermi energy is up while for the valance surface states are down. This is an indication of the quantum anomalous Hall effect where there will be a half-quantized Hall conductivity at each interface. We would like to mention that the number of QLs for Bi$_2$Se$_3$ should be 6 or more to see this condition in the current setup.

\subsection{Phase diagram for slab of magnetised TI}

\subsubsection{Band topology and phase diagram}

\par
\begin{figure}
\begin{centering}
\includegraphics[scale=0.4]{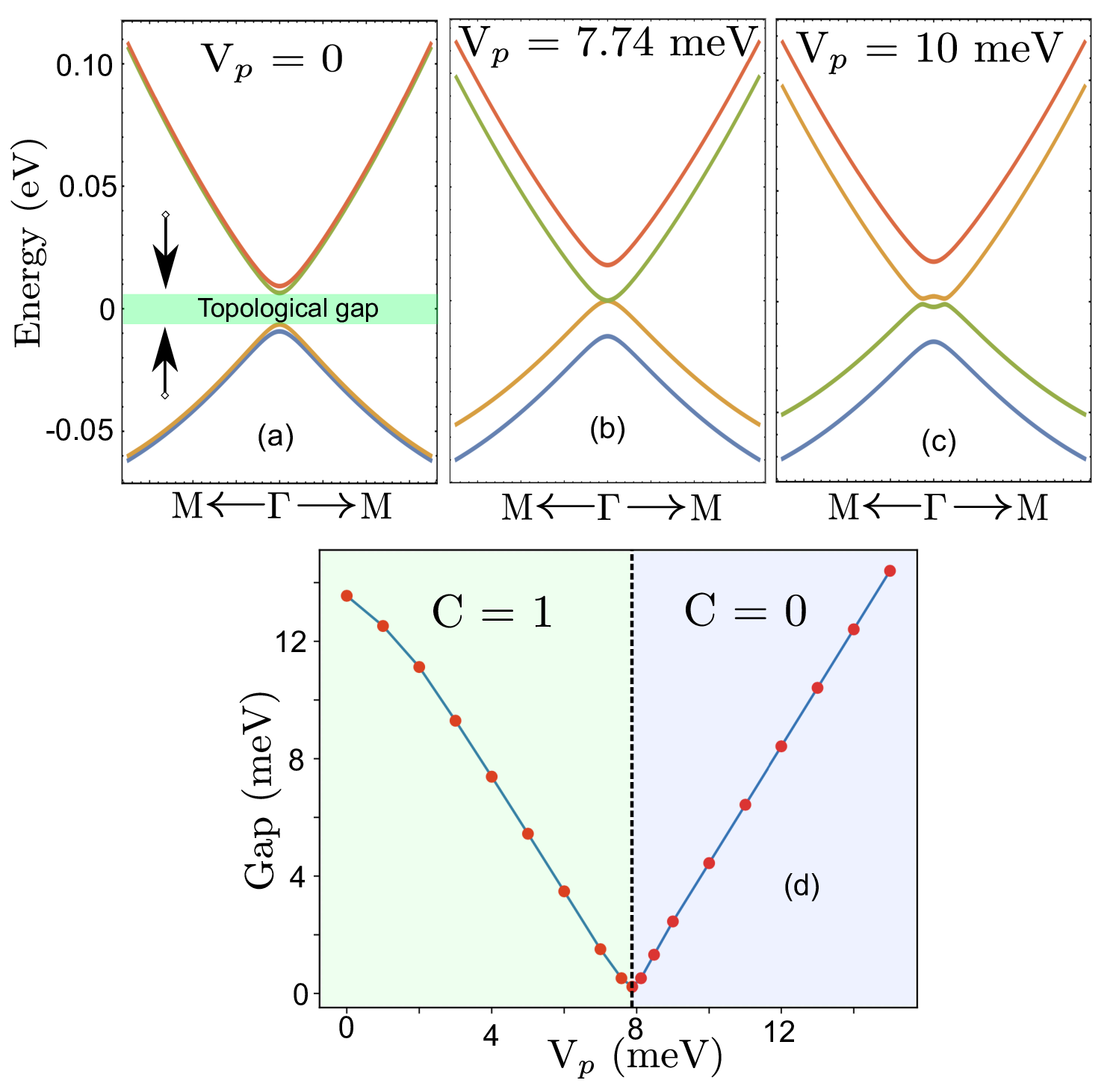}
\end{centering}
\caption{Band structure of  CrI$_3$-Bi$_2$Se$_3$-CrI$_3$ heterostructure from low energy effective four band model increasing asymmetric interfacial potential V$_p$ with (a) V$_p$ = 0 meV,  (b) V$_p$ = 7.74 meV  and (c) V$_p$ = 10 meV respectively.  (d) Dependence of topological gap and Chern numbers on V$_p$ for CrI$_3$-Bi$_2$Se$_3$-CrI$_3$ heterostructure.  A topological phase transition from topological to trivial insulator occurs at  V$_p$ = 7.74 meV.
\label{Hamil}}
\end{figure}

 Here we investigated the band topology of CrI$_3$/Bi$_2$Se$_3$/CrI$_3$ slab following the rule that i) it has Chern number $C=1$ when $\Delta^2 > M^2$ i.e in the topological phase and ii) it has Chern number $C = 0$ when $\Delta^2 < M^2$ i.e in the non-topological
\AnnaD{trivial} phase \cite{doi:10.1126/sciadv.aaw1874, doi:10.1146/annurev-conmatphys-031115-011417, doi:10.1126/science.1187485}.  Now CrI$_3$/Bi$_2$Se$_3$/CrI$_3$ slab has inversion symmetry which gives the asymmetric interfacial potential $V_p = 0$.  M depends on the thickness of the Bi$_2$Se$_3$ layers and we get a bulk band gap of $\sim$ 14 meV with M $\sim$ 1  meV.  The corresponding energy spectrum for  CrI$_3$/Bi$_2$Se$_3$/CrI$_3$ slab are presented in fig.\ref{Hamil}(a).   In this phase,  CrI$_3$/Bi$_2$Se$_3$/CrI$_3$ slab is in topological regime ($C = 1$) satisfying the condition of $\Delta^2 > M^2$.  

\par Now we slowly increase the asymmetric interfacial potential $V_p$ to explore the effect of breaking of inversion symmetry and studied the evolution of band structure of CrI$_3$/Bi$_2$Se$_3$/CrI$_3$ slab as shown in fig.\ref{Hamil}(b)-(c).  It is in the topological phase with $C = 1$ for $V_p < 7.74$ meV because of $\Delta^2 > M^2 + V_p^2$.  The topological gap at $\Gamma$ point closes with $V_p = 7.74$ meV ($\Delta^2 = M^2 + V_p^2$) which leads to in Dirac semi-metallic phase (see fig.\ref{Hamil}(b)) and then switches to a trivial phase ($C = 0$) from topological phase with further increasing the asymmetric interfacial potential $V_p > 7.74$ meV ($\Delta^2 < M^2 + V_p^2$) as shown in fig.\ref{Hamil}(c). fig.\ref{Hamil}(d) represents the corresponding phase diagram with asymmetric interfacial potential $V_p$.   CrI$_3$/Bi$_2$Se$_3$/CrI$_3$ slab is in topological insulating phase ($C=1$) with a topological gap at $\Gamma$ point for $V_p < 7.74$ meV. However,  we observe a phase transition from topological insulating phase to normal insulator ($C=0$) for $V_p > 7.74$ meV and a Dirac semi-metallic phase appears at $V_p = 7.74$ meV.

\subsection{Topology in ribbon geometry}

\begin{figure}[ht]
\includegraphics[scale=0.35]{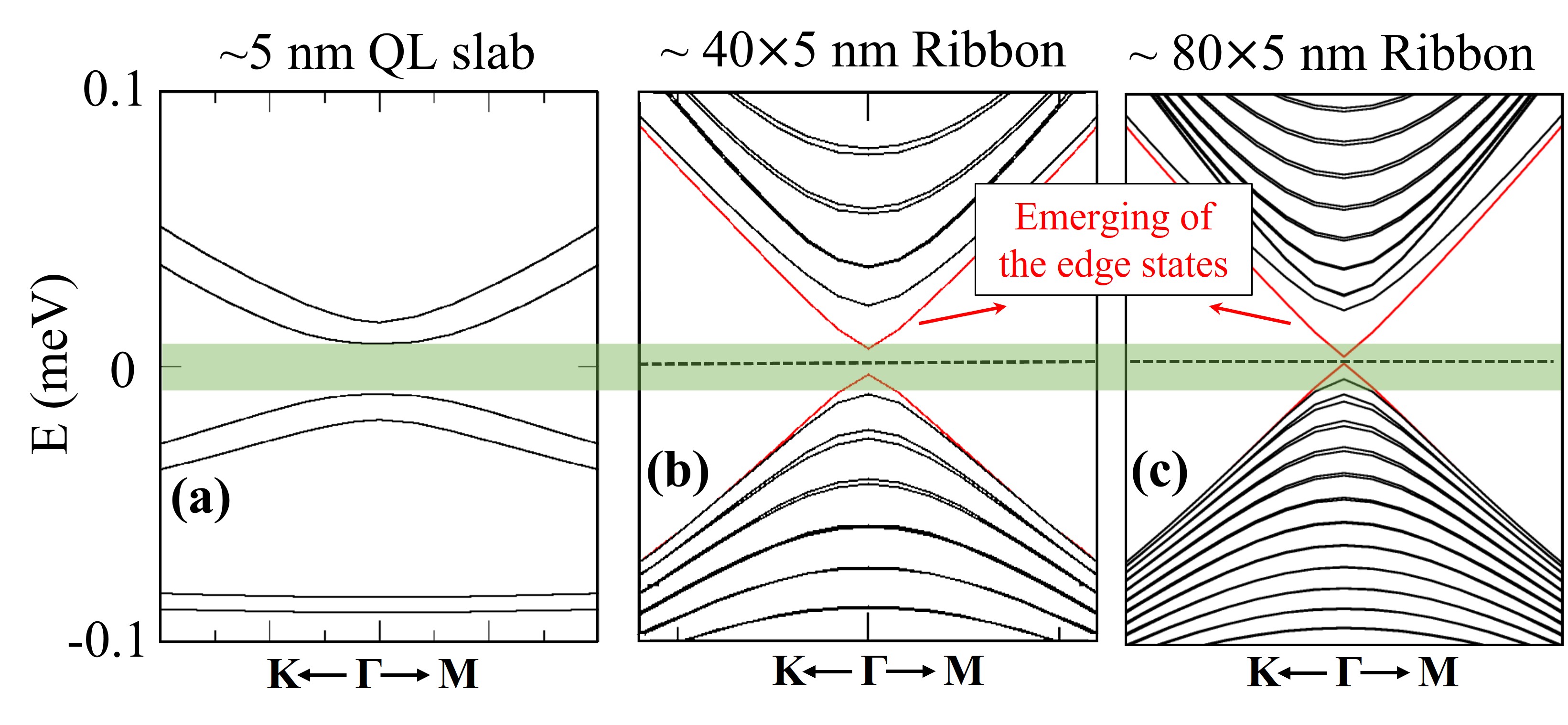}
\caption{a) The induced exchange gap at the Dirac point of the 6~QL heterostructure slab shown in Fig 1(a). b) Emerging of the edge states when we make a 40 nm and c) 80 nm ribbon of a 6~QL slab\label{Bands}}.
\end{figure}

\begin{figure} [htb]
\includegraphics[scale=0.4]{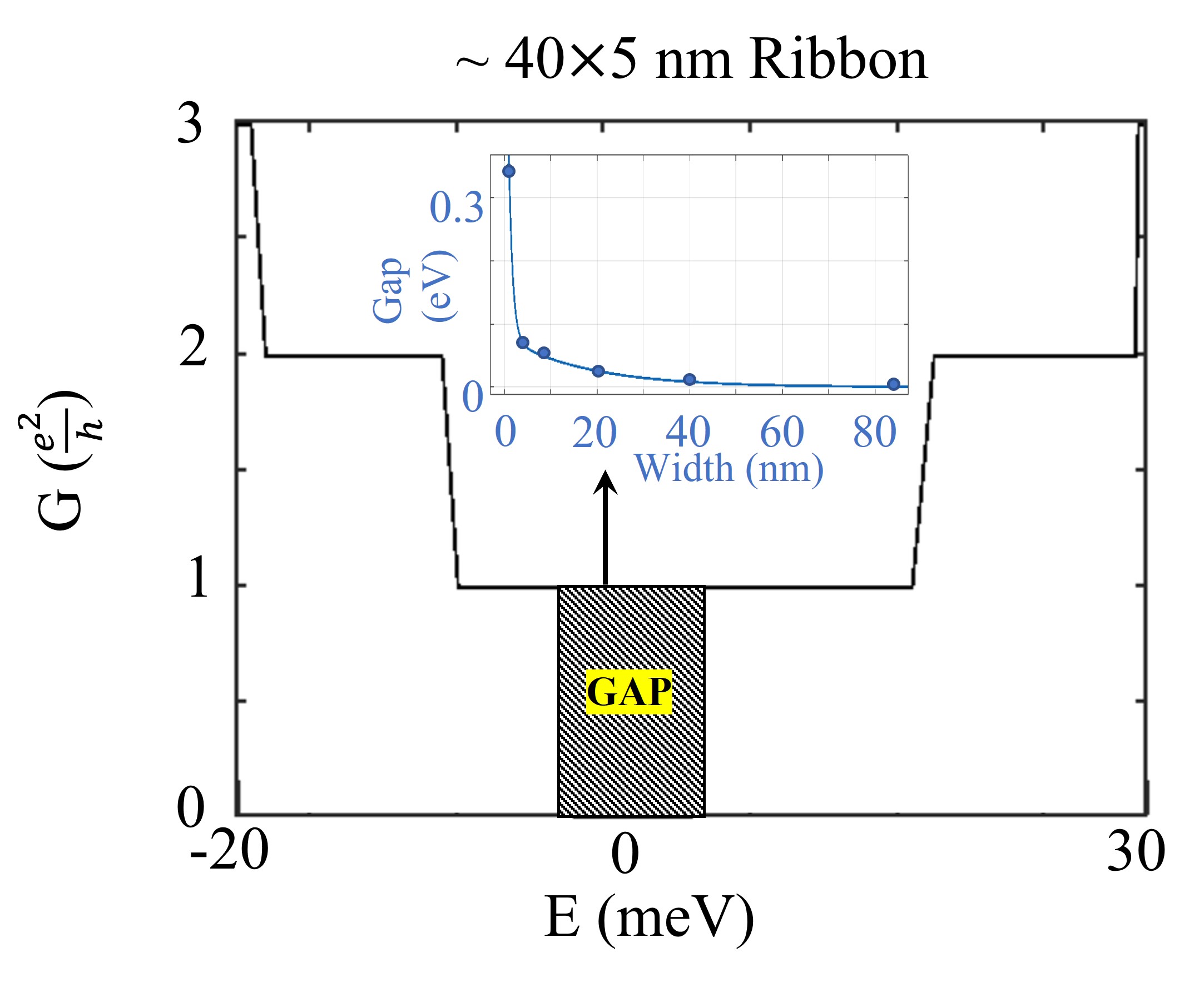}
\caption{The spin conductance for a nanoribbon with $40$~nm width and 6~QL. The inset shows that the size of the gap at the edge states exponentially reduces as the width of the ribbon is increased.
\label{Conductance}}
\end{figure}

If we consider a nanoribbon by limiting the structure also in the x direction, there should be chiral edge states inside the exchange gap. In this configuration, $k_y$ is still a good number and we can plot the band structure along this direction. As it is shown in  fig.~\ref{Bands}(a), there is an induced exchange gap for the 6QL slab, and by making a nanoribbon we can see the emergence of the edge states inside this gap in  fig.~\ref{Bands}(b). These edge states are spin-polarized and chiral which characterized the Chern phase of the system.
However, due to the coupling between the lateral side walls, the width should be larger than a critical value in order to see a perfect quantization, otherwise the edge states are still gapped. By increasing the width of the system we can see form fig.~\ref{Bands}(c) that the gap will shrink and vanish.  As it is shown in the inset of fig.~\ref{Conductance}, the size of the gap at the edge states decreases exponentially by increasing the width of the nanoribbon, and for the current setup minimum width of 80 nm gives a well-defined gapless edge states. If we compare this value with the minimum thickness along the z direction which is about 25 nm to have decoupled states and avoid opening the gap at the surface states, this values is considerably larger and the reason is that the real-space distribution of the wave functions are more broaden for the side walls compared to the top and bottom surfaces.

Given the Hamiltonian of the central channel, the spin-dependent retarded ($r$) and advanced ($a$) Green\noindent 's functions are given by
$\boldsymbol{{\cal G}}^{r} (E)=[E^{+}\boldsymbol{I}-\boldsymbol{H}-\boldsymbol{\Sigma}_{L}^{r}(E)-$ $\boldsymbol{\Sigma}_{R}^{r}(E)]^{-1}=[\boldsymbol{{\cal G}}^{a}(E)]^{\dag }$ is the self-energy due to the connection of the left (right) semi-infinite electrode to the central channel. $\boldsymbol{g}_{L(R)}$ is the surface Green's
function of the left(right) lead; $\boldsymbol{g}_{R}$ is calculated iteratively by adding one unit cell from the central region at a time using the method of Ref.~\cite{Sancho1984}.
Therefore, the Green's functions of a central region consisting of $M$ unit cells each containing $N$ atoms can be efficiently calculated using $2 N \times 2N$ matrices, where 2 comes from the spin-degree of freedom.
$\boldsymbol{H}_{L(R),C}$ is
the tunneling Hamiltonian between the central region and the left(right) lead.
The Hamiltonian of the leads, as well as the tunneling
Hamiltonian, is identical to that of the central region, namely it is given by Eq.~\ref{eq:2}, except that it does not include the exchange field.
Now that we have the Green's functions, the conductance can be calculated as
$G_{\sigma \sigma'}(E)=\frac{e^{2}}{h}\mathrm{Tr} [\boldsymbol{\Gamma}^{L}(E) \boldsymbol{{\cal G}}^{r} (E) \boldsymbol{\Gamma}^{R} (E) \boldsymbol{{\cal G}}^{a} (E)]_{\sigma \sigma'}=\frac{e^{2} }{h} T_{\sigma \sigma '}(E)$, 
where
$\boldsymbol{\Gamma}^{L,R} = i[ \boldsymbol{\Sigma}^r_{L/R}- (
\boldsymbol{\Sigma}^r_{L/R} )^\dag]$ is the broadening due to the electrode
contacts and the trace is taken over the sites in the $\sigma \sigma'$ sub-matrix.

In fig.~\ref{Conductance}, we have calculated the longitudinal conductance in the central channel. Due to the large matrices involved in this calculation, we have chosen a few value of energies to see the general behavior of the conductance. As it is shown in this figure, the conductance has a step-wise shape with the first step at $G=\frac{e^2}{h}= G_0$. We interpret this value as having two $\frac{e^2}{2h}$ corresponding to each edge considering both the top and bottom surfaces. The conductance remains constant to this quantized value when the Fermi energy varies within the energy interval $[-8,19]$ meV, however, due to the coupling of the side-walls for this width there will be a gap of $8$ meV with no edge states. This is the energy interval where only two chiral edge states are present but outside this interval, other states intervene and the conductance changes.

%% file: 04_Conclusions.tex
\section{Conclusions}
In summary, our investigation into the interplay of nontrivial band topology and magnetism in three-dimensional topological insulators has led to intriguing findings with significant implications for spintronic applications. By breaking time-reversal symmetry through the proximity effect, we have unveiled novel quantum phases, particularly the quantum anomalous Hall (QAH) phase, which showcases remarkable transport properties attributed to sidewall chiral edge states.
Our study of finite slabs within this context has revealed that the contribution of surface states depends critically on the system's cross-section and thickness. The choice of these parameters governs the extent to which the nontrivial sidewall states, connecting the top and bottom surfaces, influence the material's behavior.
Specifically, we have examined a heterostructure comprising the topological insulator Bi$_2$Se$_3$, sandwiched between two-dimensional magnetic monolayers of CrI$_3$. Employing a combination of density functional theory (DFT), tight-binding calculations, and the non-equilibrium Green’s function formalism, we have observed the emergence of a well-defined exchange gap in the band structure, accommodating spin-polarized edge states with distinctive transport characteristics.
Furthermore, our investigation extends to the study of width and finite-size effects on the transmission and topological properties of this magnetized topological insulator nanoribbon. These findings contribute valuable insights into the design and understanding of materials with tailored topological and transport properties, holding great promise for future advancements in the field of spintronics and quantum information technologies.

%% file: main.bbl
\begin{thebibliography}{39}
\expandafter\ifx\csname natexlab\endcsname\relax\def\natexlab#1{#1}\fi
\expandafter\ifx\csname bibnamefont\endcsname\relax
  \def\bibnamefont#1{#1}\fi
\expandafter\ifx\csname bibfnamefont\endcsname\relax
  \def\bibfnamefont#1{#1}\fi
\expandafter\ifx\csname citenamefont\endcsname\relax
  \def\citenamefont#1{#1}\fi
\expandafter\ifx\csname url\endcsname\relax
  \def\url#1{\texttt{#1}}\fi
\expandafter\ifx\csname urlprefix\endcsname\relax\def\urlprefix{URL }\fi
\providecommand{\bibinfo}[2]{#2}
\providecommand{\eprint}[2][]{\url{#2}}

\bibitem[{\citenamefont{Hasan and Kane}(2010)}]{hasan2010}
\bibinfo{author}{\bibfnamefont{M.~Z.} \bibnamefont{Hasan}} \bibnamefont{and}
  \bibinfo{author}{\bibfnamefont{C.~L.} \bibnamefont{Kane}},
  \bibinfo{journal}{Rev. Mod. Phys.} \textbf{\bibinfo{volume}{82}},
  \bibinfo{pages}{3045} (\bibinfo{year}{2010}),
  \urlprefix\url{http://link.aps.org/doi/10.1103/RevModPhys.82.3045}.

\bibitem[{\citenamefont{Qi and Zhang}(2011)}]{XLQi}
\bibinfo{author}{\bibfnamefont{X.-L.} \bibnamefont{Qi}} \bibnamefont{and}
  \bibinfo{author}{\bibfnamefont{S.-C.} \bibnamefont{Zhang}},
  \bibinfo{journal}{Rev. Mod. Phys.} \textbf{\bibinfo{volume}{83}},
  \bibinfo{pages}{1057} (\bibinfo{year}{2011}).

\bibitem[{\citenamefont{Ando}(2013)}]{Ando2013}
\bibinfo{author}{\bibfnamefont{Y.}~\bibnamefont{Ando}}, \bibinfo{journal}{J.
  Phys. Soc. Jpn.} \textbf{\bibinfo{volume}{82}}, \bibinfo{pages}{102001}
  (\bibinfo{year}{2013}).

\bibitem[{\citenamefont{Bansil et~al.}(2016)\citenamefont{Bansil, Lin, and
  Das}}]{Bansil2016}
\bibinfo{author}{\bibfnamefont{A.}~\bibnamefont{Bansil}},
  \bibinfo{author}{\bibfnamefont{H.}~\bibnamefont{Lin}}, \bibnamefont{and}
  \bibinfo{author}{\bibfnamefont{T.}~\bibnamefont{Das}}, \bibinfo{journal}{Rev.
  Mod. Phys.} \textbf{\bibinfo{volume}{88}}, \bibinfo{pages}{021004}
  (\bibinfo{year}{2016}),
  \urlprefix\url{https://link.aps.org/doi/10.1103/RevModPhys.88.021004}.

\bibitem[{\citenamefont{Fu and Kane}(2007)}]{Fu2007}
\bibinfo{author}{\bibfnamefont{L.}~\bibnamefont{Fu}} \bibnamefont{and}
  \bibinfo{author}{\bibfnamefont{C.~L.} \bibnamefont{Kane}},
  \bibinfo{journal}{Phys. Rev. B} \textbf{\bibinfo{volume}{76}},
  \bibinfo{pages}{045302} (\bibinfo{year}{2007}).

\bibitem[{\citenamefont{Qi et~al.}(2008)\citenamefont{Qi, Hughes, and
  Zhang}}]{Qi2008}
\bibinfo{author}{\bibfnamefont{X.-L.} \bibnamefont{Qi}},
  \bibinfo{author}{\bibfnamefont{T.~L.} \bibnamefont{Hughes}},
  \bibnamefont{and} \bibinfo{author}{\bibfnamefont{S.-C.} \bibnamefont{Zhang}},
  \bibinfo{journal}{Phys. Rev. B} \textbf{\bibinfo{volume}{78}},
  \bibinfo{pages}{195424} (\bibinfo{year}{2008}).

\bibitem[{\citenamefont{Essin et~al.}(2009)\citenamefont{Essin, Moore, and
  Vanderbilt}}]{Essin2009}
\bibinfo{author}{\bibfnamefont{A.~M.} \bibnamefont{Essin}},
  \bibinfo{author}{\bibfnamefont{J.~E.} \bibnamefont{Moore}}, \bibnamefont{and}
  \bibinfo{author}{\bibfnamefont{D.}~\bibnamefont{Vanderbilt}},
  \bibinfo{journal}{Phys. Rev. Lett.} \textbf{\bibinfo{volume}{102}},
  \bibinfo{pages}{146805} (\bibinfo{year}{2009}),
  \urlprefix\url{https://link.aps.org/doi/10.1103/PhysRevLett.102.146805}.

\bibitem[{\citenamefont{Vanderbilt}(2018)}]{Vanderbilt}
\bibinfo{author}{\bibfnamefont{D.}~\bibnamefont{Vanderbilt}},
  \emph{\bibinfo{title}{Berry Phases in Electronic Structure Theory}}
  (\bibinfo{publisher}{Cambridge University Press}, \bibinfo{year}{2018}).

\bibitem[{\citenamefont{Yu et~al.}(2010{\natexlab{a}})\citenamefont{Yu, Zhang,
  Zhang, Zhang, Dai, and Fang}}]{Yu2010}
\bibinfo{author}{\bibfnamefont{R.}~\bibnamefont{Yu}},
  \bibinfo{author}{\bibfnamefont{W.}~\bibnamefont{Zhang}},
  \bibinfo{author}{\bibfnamefont{H.-J.} \bibnamefont{Zhang}},
  \bibinfo{author}{\bibfnamefont{S.-C.} \bibnamefont{Zhang}},
  \bibinfo{author}{\bibfnamefont{X.}~\bibnamefont{Dai}}, \bibnamefont{and}
  \bibinfo{author}{\bibfnamefont{Z.}~\bibnamefont{Fang}},
  \bibinfo{journal}{Science} \textbf{\bibinfo{volume}{329}},
  \bibinfo{pages}{61} (\bibinfo{year}{2010}{\natexlab{a}}).

\bibitem[{\citenamefont{Chang et~al.}(2013)\citenamefont{Chang, Zhang, Feng,
  Shen, Zhang, Guo, Li, Ou, Wei, Wang et~al.}}]{Chang2013}
\bibinfo{author}{\bibfnamefont{C.-Z.} \bibnamefont{Chang}},
  \bibinfo{author}{\bibfnamefont{J.}~\bibnamefont{Zhang}},
  \bibinfo{author}{\bibfnamefont{X.}~\bibnamefont{Feng}},
  \bibinfo{author}{\bibfnamefont{J.}~\bibnamefont{Shen}},
  \bibinfo{author}{\bibfnamefont{Z.}~\bibnamefont{Zhang}},
  \bibinfo{author}{\bibfnamefont{M.}~\bibnamefont{Guo}},
  \bibinfo{author}{\bibfnamefont{K.}~\bibnamefont{Li}},
  \bibinfo{author}{\bibfnamefont{Y.}~\bibnamefont{Ou}},
  \bibinfo{author}{\bibfnamefont{P.}~\bibnamefont{Wei}},
  \bibinfo{author}{\bibfnamefont{L.-L.} \bibnamefont{Wang}},
  \bibnamefont{et~al.}, \bibinfo{journal}{Science}
  \textbf{\bibinfo{volume}{340}}, \bibinfo{pages}{167} (\bibinfo{year}{2013}),
  ISSN \bibinfo{issn}{0036-8075}.

\bibitem[{\citenamefont{Liu et~al.}(2016{\natexlab{a}})\citenamefont{Liu,
  Zhang, and Qi}}]{Liu2016}
\bibinfo{author}{\bibfnamefont{C.-X.} \bibnamefont{Liu}},
  \bibinfo{author}{\bibfnamefont{S.-C.} \bibnamefont{Zhang}}, \bibnamefont{and}
  \bibinfo{author}{\bibfnamefont{X.-L.} \bibnamefont{Qi}},
  \bibinfo{journal}{Annual Review of Condensed Matter Physics}
  \textbf{\bibinfo{volume}{7}}, \bibinfo{pages}{301}
  (\bibinfo{year}{2016}{\natexlab{a}}).

\bibitem[{\citenamefont{Chang and Li}(2016)}]{Chang2016}
\bibinfo{author}{\bibfnamefont{C.-Z.} \bibnamefont{Chang}} \bibnamefont{and}
  \bibinfo{author}{\bibfnamefont{M.}~\bibnamefont{Li}},
  \bibinfo{journal}{Journal of Physics: Condensed Matter}
  \textbf{\bibinfo{volume}{28}}, \bibinfo{pages}{123002}
  (\bibinfo{year}{2016}).

\bibitem[{\citenamefont{Luo and Qi}(2013)}]{Luo2013}
\bibinfo{author}{\bibfnamefont{W.}~\bibnamefont{Luo}} \bibnamefont{and}
  \bibinfo{author}{\bibfnamefont{X.-L.} \bibnamefont{Qi}},
  \bibinfo{journal}{Phys. Rev. B} \textbf{\bibinfo{volume}{87}},
  \bibinfo{pages}{085431} (\bibinfo{year}{2013}),
  \urlprefix\url{https://link.aps.org/doi/10.1103/PhysRevB.87.085431}.

\bibitem[{\citenamefont{Eremeev et~al.}(2013)\citenamefont{Eremeev, Men'shov,
  Tugushev, Echenique, and Chulkov}}]{Eremeev2013}
\bibinfo{author}{\bibfnamefont{S.~V.} \bibnamefont{Eremeev}},
  \bibinfo{author}{\bibfnamefont{V.~N.} \bibnamefont{Men'shov}},
  \bibinfo{author}{\bibfnamefont{V.~V.} \bibnamefont{Tugushev}},
  \bibinfo{author}{\bibfnamefont{P.~M.} \bibnamefont{Echenique}},
  \bibnamefont{and} \bibinfo{author}{\bibfnamefont{E.~V.}
  \bibnamefont{Chulkov}}, \bibinfo{journal}{Phys. Rev. B}
  \textbf{\bibinfo{volume}{88}}, \bibinfo{pages}{144430}
  (\bibinfo{year}{2013}),
  \urlprefix\url{https://link.aps.org/doi/10.1103/PhysRevB.88.144430}.

\bibitem[{\citenamefont{Wei et~al.}(2013)\citenamefont{Wei, Katmis, Assaf,
  Steinberg, Jarillo-Herrero, Heiman, and Moodera}}]{Wei2013}
\bibinfo{author}{\bibfnamefont{P.}~\bibnamefont{Wei}},
  \bibinfo{author}{\bibfnamefont{F.}~\bibnamefont{Katmis}},
  \bibinfo{author}{\bibfnamefont{B.~A.} \bibnamefont{Assaf}},
  \bibinfo{author}{\bibfnamefont{H.}~\bibnamefont{Steinberg}},
  \bibinfo{author}{\bibfnamefont{P.}~\bibnamefont{Jarillo-Herrero}},
  \bibinfo{author}{\bibfnamefont{D.}~\bibnamefont{Heiman}}, \bibnamefont{and}
  \bibinfo{author}{\bibfnamefont{J.~S.} \bibnamefont{Moodera}},
  \bibinfo{journal}{Phys. Rev. Lett.} \textbf{\bibinfo{volume}{110}},
  \bibinfo{pages}{186807} (\bibinfo{year}{2013}).

\bibitem[{\citenamefont{Yang et~al.}(2013)\citenamefont{Yang, Dolev, Zhang,
  Zhao, Fried, Schemm, Liu, Palevski, Marshall, Risbud et~al.}}]{Yang2013}
\bibinfo{author}{\bibfnamefont{Q.~I.} \bibnamefont{Yang}},
  \bibinfo{author}{\bibfnamefont{M.}~\bibnamefont{Dolev}},
  \bibinfo{author}{\bibfnamefont{L.}~\bibnamefont{Zhang}},
  \bibinfo{author}{\bibfnamefont{J.}~\bibnamefont{Zhao}},
  \bibinfo{author}{\bibfnamefont{A.~D.} \bibnamefont{Fried}},
  \bibinfo{author}{\bibfnamefont{E.}~\bibnamefont{Schemm}},
  \bibinfo{author}{\bibfnamefont{M.}~\bibnamefont{Liu}},
  \bibinfo{author}{\bibfnamefont{A.}~\bibnamefont{Palevski}},
  \bibinfo{author}{\bibfnamefont{A.~F.} \bibnamefont{Marshall}},
  \bibinfo{author}{\bibfnamefont{S.~H.} \bibnamefont{Risbud}},
  \bibnamefont{et~al.}, \bibinfo{journal}{Phys. Rev. B(R)}
  \textbf{\bibinfo{volume}{88}}, \bibinfo{pages}{081407}
  (\bibinfo{year}{2013}).

\bibitem[{\citenamefont{Lang et~al.}(2014)\citenamefont{Lang, Montazeri,
  Onbasli, Kou, Fan, Upadhyaya, Yao, Liu, Jiang, Jiang et~al.}}]{Lang2014}
\bibinfo{author}{\bibfnamefont{M.}~\bibnamefont{Lang}},
  \bibinfo{author}{\bibfnamefont{M.}~\bibnamefont{Montazeri}},
  \bibinfo{author}{\bibfnamefont{M.~C.} \bibnamefont{Onbasli}},
  \bibinfo{author}{\bibfnamefont{X.}~\bibnamefont{Kou}},
  \bibinfo{author}{\bibfnamefont{Y.}~\bibnamefont{Fan}},
  \bibinfo{author}{\bibfnamefont{P.}~\bibnamefont{Upadhyaya}},
  \bibinfo{author}{\bibfnamefont{K.}~\bibnamefont{Yao}},
  \bibinfo{author}{\bibfnamefont{F.}~\bibnamefont{Liu}},
  \bibinfo{author}{\bibfnamefont{Y.}~\bibnamefont{Jiang}},
  \bibinfo{author}{\bibfnamefont{W.}~\bibnamefont{Jiang}},
  \bibnamefont{et~al.}, \bibinfo{journal}{Nano Lett.}
  \textbf{\bibinfo{volume}{14}}, \bibinfo{pages}{3459} (\bibinfo{year}{2014}).

\bibitem[{\citenamefont{Katmis et~al.}(2016)\citenamefont{Katmis, Lauter,
  Nogueira, Assaf, Jamer, Wei, Satpati, Freeland, Eremin, Heiman
  et~al.}}]{Katmis2016}
\bibinfo{author}{\bibfnamefont{F.}~\bibnamefont{Katmis}},
  \bibinfo{author}{\bibfnamefont{V.}~\bibnamefont{Lauter}},
  \bibinfo{author}{\bibfnamefont{F.~S.} \bibnamefont{Nogueira}},
  \bibinfo{author}{\bibfnamefont{B.~A.} \bibnamefont{Assaf}},
  \bibinfo{author}{\bibfnamefont{M.~E.} \bibnamefont{Jamer}},
  \bibinfo{author}{\bibfnamefont{P.}~\bibnamefont{Wei}},
  \bibinfo{author}{\bibfnamefont{B.}~\bibnamefont{Satpati}},
  \bibinfo{author}{\bibfnamefont{J.~W.} \bibnamefont{Freeland}},
  \bibinfo{author}{\bibfnamefont{I.}~\bibnamefont{Eremin}},
  \bibinfo{author}{\bibfnamefont{D.}~\bibnamefont{Heiman}},
  \bibnamefont{et~al.}, \bibinfo{journal}{Nature(London)}
  \textbf{\bibinfo{volume}{533}}, \bibinfo{pages}{513} (\bibinfo{year}{2016}).

\bibitem[{\citenamefont{Tang et~al.}(2017)\citenamefont{Tang, Chang, Zhao, Liu,
  Jiang, Liu, McCartney, Smith, Chen, Moodera et~al.}}]{Tang2017}
\bibinfo{author}{\bibfnamefont{C.}~\bibnamefont{Tang}},
  \bibinfo{author}{\bibfnamefont{C.-Z.} \bibnamefont{Chang}},
  \bibinfo{author}{\bibfnamefont{G.}~\bibnamefont{Zhao}},
  \bibinfo{author}{\bibfnamefont{Y.}~\bibnamefont{Liu}},
  \bibinfo{author}{\bibfnamefont{Z.}~\bibnamefont{Jiang}},
  \bibinfo{author}{\bibfnamefont{C.-X.} \bibnamefont{Liu}},
  \bibinfo{author}{\bibfnamefont{M.~R.} \bibnamefont{McCartney}},
  \bibinfo{author}{\bibfnamefont{D.~J.} \bibnamefont{Smith}},
  \bibinfo{author}{\bibfnamefont{T.}~\bibnamefont{Chen}},
  \bibinfo{author}{\bibfnamefont{J.~S.} \bibnamefont{Moodera}},
  \bibnamefont{et~al.}, \bibinfo{journal}{Sci. Adv.}
  \textbf{\bibinfo{volume}{3}}, \bibinfo{pages}{e1700307}
  (\bibinfo{year}{2017}).

\bibitem[{\citenamefont{Hirahara et~al.}(2017)\citenamefont{Hirahara, Eremeev,
  Shirasawa, Okuyama, Kubo, Nakanishi, Akiyama, Takayama, Hajiri, Ideta
  et~al.}}]{Hirahara2017}
\bibinfo{author}{\bibfnamefont{T.}~\bibnamefont{Hirahara}},
  \bibinfo{author}{\bibfnamefont{S.~V.} \bibnamefont{Eremeev}},
  \bibinfo{author}{\bibfnamefont{T.}~\bibnamefont{Shirasawa}},
  \bibinfo{author}{\bibfnamefont{Y.}~\bibnamefont{Okuyama}},
  \bibinfo{author}{\bibfnamefont{T.}~\bibnamefont{Kubo}},
  \bibinfo{author}{\bibfnamefont{R.}~\bibnamefont{Nakanishi}},
  \bibinfo{author}{\bibfnamefont{R.}~\bibnamefont{Akiyama}},
  \bibinfo{author}{\bibfnamefont{A.}~\bibnamefont{Takayama}},
  \bibinfo{author}{\bibfnamefont{T.}~\bibnamefont{Hajiri}},
  \bibinfo{author}{\bibfnamefont{S.}~\bibnamefont{Ideta}},
  \bibnamefont{et~al.}, \bibinfo{journal}{Nano Lett.}
  \textbf{\bibinfo{volume}{17}}, \bibinfo{pages}{3493} (\bibinfo{year}{2017}).

\bibitem[{\citenamefont{Zhu et~al.}(2018)\citenamefont{Zhu, Meng, Liang, Shi,
  Zhao, Cheng, Li, Zhai, Lu, Chen et~al.}}]{Zhu2018}
\bibinfo{author}{\bibfnamefont{S.}~\bibnamefont{Zhu}},
  \bibinfo{author}{\bibfnamefont{D.}~\bibnamefont{Meng}},
  \bibinfo{author}{\bibfnamefont{G.}~\bibnamefont{Liang}},
  \bibinfo{author}{\bibfnamefont{G.}~\bibnamefont{Shi}},
  \bibinfo{author}{\bibfnamefont{P.}~\bibnamefont{Zhao}},
  \bibinfo{author}{\bibfnamefont{P.}~\bibnamefont{Cheng}},
  \bibinfo{author}{\bibfnamefont{Y.}~\bibnamefont{Li}},
  \bibinfo{author}{\bibfnamefont{X.}~\bibnamefont{Zhai}},
  \bibinfo{author}{\bibfnamefont{Y.}~\bibnamefont{Lu}},
  \bibinfo{author}{\bibfnamefont{L.}~\bibnamefont{Chen}}, \bibnamefont{et~al.},
  \bibinfo{journal}{Nanoscale} \textbf{\bibinfo{volume}{10}},
  \bibinfo{pages}{10041} (\bibinfo{year}{2018}).

\bibitem[{\citenamefont{Mogi et~al.}(2019)\citenamefont{Mogi, Nakajima, Ukleev,
  Tsukazaki, Yoshimi, Kawamura, Takahashi, Hanashima, Kakurai, Arima
  et~al.}}]{Mogi2019a}
\bibinfo{author}{\bibfnamefont{M.}~\bibnamefont{Mogi}},
  \bibinfo{author}{\bibfnamefont{T.}~\bibnamefont{Nakajima}},
  \bibinfo{author}{\bibfnamefont{V.}~\bibnamefont{Ukleev}},
  \bibinfo{author}{\bibfnamefont{A.}~\bibnamefont{Tsukazaki}},
  \bibinfo{author}{\bibfnamefont{R.}~\bibnamefont{Yoshimi}},
  \bibinfo{author}{\bibfnamefont{M.}~\bibnamefont{Kawamura}},
  \bibinfo{author}{\bibfnamefont{K.~S.} \bibnamefont{Takahashi}},
  \bibinfo{author}{\bibfnamefont{T.}~\bibnamefont{Hanashima}},
  \bibinfo{author}{\bibfnamefont{K.}~\bibnamefont{Kakurai}},
  \bibinfo{author}{\bibfnamefont{T.}~\bibnamefont{Arima}},
  \bibnamefont{et~al.}, \bibinfo{journal}{Phys. Rev. Lett.}
  \textbf{\bibinfo{volume}{123}}, \bibinfo{pages}{016804}
  (\bibinfo{year}{2019}).

\bibitem[{\citenamefont{He et~al.}(2018)\citenamefont{He, Yin, Yu, Grutter,
  Pan, Chen, Che, Yu, Zhang, Shao et~al.}}]{he_PRL2018}
\bibinfo{author}{\bibfnamefont{Q.~L.} \bibnamefont{He}},
  \bibinfo{author}{\bibfnamefont{G.}~\bibnamefont{Yin}},
  \bibinfo{author}{\bibfnamefont{L.}~\bibnamefont{Yu}},
  \bibinfo{author}{\bibfnamefont{A.~J.} \bibnamefont{Grutter}},
  \bibinfo{author}{\bibfnamefont{L.}~\bibnamefont{Pan}},
  \bibinfo{author}{\bibfnamefont{C.-Z.} \bibnamefont{Chen}},
  \bibinfo{author}{\bibfnamefont{X.}~\bibnamefont{Che}},
  \bibinfo{author}{\bibfnamefont{G.}~\bibnamefont{Yu}},
  \bibinfo{author}{\bibfnamefont{B.}~\bibnamefont{Zhang}},
  \bibinfo{author}{\bibfnamefont{Q.}~\bibnamefont{Shao}}, \bibnamefont{et~al.},
  \bibinfo{journal}{Phys. Rev. Lett.} \textbf{\bibinfo{volume}{121}},
  \bibinfo{pages}{096802} (\bibinfo{year}{2018}),
  \urlprefix\url{https://doi.org/10.1103/PhysRevLett.121.096802}.

\bibitem[{\citenamefont{Yang et~al.}(2020)\citenamefont{Yang, Pan, Grutter,
  Wang, Che, He, Wu, Gilbert, Shafer, Arenholz et~al.}}]{Yang2020}
\bibinfo{author}{\bibfnamefont{C.-Y.} \bibnamefont{Yang}},
  \bibinfo{author}{\bibfnamefont{L.}~\bibnamefont{Pan}},
  \bibinfo{author}{\bibfnamefont{A.~J.} \bibnamefont{Grutter}},
  \bibinfo{author}{\bibfnamefont{H.}~\bibnamefont{Wang}},
  \bibinfo{author}{\bibfnamefont{X.}~\bibnamefont{Che}},
  \bibinfo{author}{\bibfnamefont{Q.~L.} \bibnamefont{He}},
  \bibinfo{author}{\bibfnamefont{Y.}~\bibnamefont{Wu}},
  \bibinfo{author}{\bibfnamefont{D.~A.} \bibnamefont{Gilbert}},
  \bibinfo{author}{\bibfnamefont{P.}~\bibnamefont{Shafer}},
  \bibinfo{author}{\bibfnamefont{E.}~\bibnamefont{Arenholz}},
  \bibnamefont{et~al.}, \bibinfo{journal}{Sci. Adv.}
  \textbf{\bibinfo{volume}{6}}, \bibinfo{pages}{eaaz8463}
  (\bibinfo{year}{2020}).

\bibitem[{\citenamefont{Bonbien et~al.}(2021)\citenamefont{Bonbien, Zhuo,
  Salimath, Ly, Abbout, and Manchon}}]{2102.01632}
\bibinfo{author}{\bibfnamefont{V.}~\bibnamefont{Bonbien}},
  \bibinfo{author}{\bibfnamefont{F.}~\bibnamefont{Zhuo}},
  \bibinfo{author}{\bibfnamefont{A.}~\bibnamefont{Salimath}},
  \bibinfo{author}{\bibfnamefont{O.}~\bibnamefont{Ly}},
  \bibinfo{author}{\bibfnamefont{A.}~\bibnamefont{Abbout}}, \bibnamefont{and}
  \bibinfo{author}{\bibfnamefont{A.}~\bibnamefont{Manchon}},
  \emph{\bibinfo{title}{Topological aspects of antiferromagnets}}
  (\bibinfo{year}{2021}), \eprint{arXiv:2102.01632}.

\bibitem[{\citenamefont{Watanabe et~al.}(2019)\citenamefont{Watanabe, Yoshimi,
  Kawamura, Mogi, Tsukazaki, Yu, Kawasaki, and Tokura}}]{Mogi2019b}
\bibinfo{author}{\bibfnamefont{R.}~\bibnamefont{Watanabe}},
  \bibinfo{author}{\bibfnamefont{R.}~\bibnamefont{Yoshimi}},
  \bibinfo{author}{\bibfnamefont{M.}~\bibnamefont{Kawamura}},
  \bibinfo{author}{\bibfnamefont{M.}~\bibnamefont{Mogi}},
  \bibinfo{author}{\bibfnamefont{A.}~\bibnamefont{Tsukazaki}},
  \bibinfo{author}{\bibfnamefont{X.~Z.} \bibnamefont{Yu}},
  \bibinfo{author}{\bibfnamefont{M.}~\bibnamefont{Kawasaki}}, \bibnamefont{and}
  \bibinfo{author}{\bibfnamefont{Y.}~\bibnamefont{Tokura}},
  \bibinfo{journal}{Appl. Phys. Lett.} \textbf{\bibinfo{volume}{115}},
  \bibinfo{pages}{102403} (\bibinfo{year}{2019}).

\bibitem[{\citenamefont{Huang et~al.}(2017)\citenamefont{Huang, Clark,
  Navarro-Moratalla, Klein, Cheng, Seyler, Zhong, Schmidgall, McGuire, Cobden
  et~al.}}]{Huang2017}
\bibinfo{author}{\bibfnamefont{B.}~\bibnamefont{Huang}},
  \bibinfo{author}{\bibfnamefont{G.}~\bibnamefont{Clark}},
  \bibinfo{author}{\bibfnamefont{E.}~\bibnamefont{Navarro-Moratalla}},
  \bibinfo{author}{\bibfnamefont{D.~R.} \bibnamefont{Klein}},
  \bibinfo{author}{\bibfnamefont{R.}~\bibnamefont{Cheng}},
  \bibinfo{author}{\bibfnamefont{K.~L.} \bibnamefont{Seyler}},
  \bibinfo{author}{\bibfnamefont{D.}~\bibnamefont{Zhong}},
  \bibinfo{author}{\bibfnamefont{E.}~\bibnamefont{Schmidgall}},
  \bibinfo{author}{\bibfnamefont{M.~A.} \bibnamefont{McGuire}},
  \bibinfo{author}{\bibfnamefont{D.~H.} \bibnamefont{Cobden}},
  \bibnamefont{et~al.}, \bibinfo{journal}{Nature}
  \textbf{\bibinfo{volume}{546}}, \bibinfo{pages}{270} (\bibinfo{year}{2017}).

\bibitem[{\citenamefont{Hou et~al.}(2019{\natexlab{a}})\citenamefont{Hou, Kim,
  and Wu}}]{Hou2019}
\bibinfo{author}{\bibfnamefont{Y.}~\bibnamefont{Hou}},
  \bibinfo{author}{\bibfnamefont{J.}~\bibnamefont{Kim}}, \bibnamefont{and}
  \bibinfo{author}{\bibfnamefont{R.}~\bibnamefont{Wu}}, \bibinfo{journal}{Sci.
  Adv.} \textbf{\bibinfo{volume}{5}}, \bibinfo{pages}{eaaw1874}
  (\bibinfo{year}{2019}{\natexlab{a}}).

\bibitem[{\citenamefont{Kresse and Hafner}(1993)}]{vasp1}
\bibinfo{author}{\bibfnamefont{G.}~\bibnamefont{Kresse}} \bibnamefont{and}
  \bibinfo{author}{\bibfnamefont{J.}~\bibnamefont{Hafner}},
  \bibinfo{journal}{Phys. Rev. B} \textbf{\bibinfo{volume}{47}},
  \bibinfo{pages}{558} (\bibinfo{year}{1993}),
  \urlprefix\url{https://link.aps.org/doi/10.1103/PhysRevB.47.558}.

\bibitem[{\citenamefont{Kresse and Furthmüller}(1996)}]{vasp2}
\bibinfo{author}{\bibfnamefont{G.}~\bibnamefont{Kresse}} \bibnamefont{and}
  \bibinfo{author}{\bibfnamefont{J.}~\bibnamefont{Furthmüller}},
  \bibinfo{journal}{Computational Materials Science}
  \textbf{\bibinfo{volume}{6}}, \bibinfo{pages}{15 } (\bibinfo{year}{1996}),
  ISSN \bibinfo{issn}{0927-0256},
  \urlprefix\url{http://www.sciencedirect.com/science/article/pii/0927025696000080}.

\bibitem[{\citenamefont{Perdew et~al.}(1996)\citenamefont{Perdew, Burke, and
  Ernzerhof}}]{Perdew1996}
\bibinfo{author}{\bibfnamefont{J.~P.} \bibnamefont{Perdew}},
  \bibinfo{author}{\bibfnamefont{K.}~\bibnamefont{Burke}}, \bibnamefont{and}
  \bibinfo{author}{\bibfnamefont{M.}~\bibnamefont{Ernzerhof}},
  \bibinfo{journal}{Phys. Rev. Lett.} \textbf{\bibinfo{volume}{77}},
  \bibinfo{pages}{3865} (\bibinfo{year}{1996}).

\bibitem[{\citenamefont{Hou et~al.}(2019{\natexlab{b}})\citenamefont{Hou, Kim,
  and Wu}}]{doi:10.1126/sciadv.aaw1874}
\bibinfo{author}{\bibfnamefont{Y.}~\bibnamefont{Hou}},
  \bibinfo{author}{\bibfnamefont{J.}~\bibnamefont{Kim}}, \bibnamefont{and}
  \bibinfo{author}{\bibfnamefont{R.}~\bibnamefont{Wu}},
  \bibinfo{journal}{Science Advances} \textbf{\bibinfo{volume}{5}},
  \bibinfo{pages}{eaaw1874} (\bibinfo{year}{2019}{\natexlab{b}}),
  \eprint{https://www.science.org/doi/pdf/10.1126/sciadv.aaw1874},
  \urlprefix\url{https://www.science.org/doi/abs/10.1126/sciadv.aaw1874}.

\bibitem[{\citenamefont{Liu et~al.}(2016{\natexlab{b}})\citenamefont{Liu,
  Zhang, and Qi}}]{doi:10.1146/annurev-conmatphys-031115-011417}
\bibinfo{author}{\bibfnamefont{C.-X.} \bibnamefont{Liu}},
  \bibinfo{author}{\bibfnamefont{S.-C.} \bibnamefont{Zhang}}, \bibnamefont{and}
  \bibinfo{author}{\bibfnamefont{X.-L.} \bibnamefont{Qi}},
  \bibinfo{journal}{Annual Review of Condensed Matter Physics}
  \textbf{\bibinfo{volume}{7}}, \bibinfo{pages}{301}
  (\bibinfo{year}{2016}{\natexlab{b}}),
  \eprint{https://doi.org/10.1146/annurev-conmatphys-031115-011417},
  \urlprefix\url{https://doi.org/10.1146/annurev-conmatphys-031115-011417}.

\bibitem[{\citenamefont{Yu et~al.}(2010{\natexlab{b}})\citenamefont{Yu, Zhang,
  Zhang, Zhang, Dai, and Fang}}]{doi:10.1126/science.1187485}
\bibinfo{author}{\bibfnamefont{R.}~\bibnamefont{Yu}},
  \bibinfo{author}{\bibfnamefont{W.}~\bibnamefont{Zhang}},
  \bibinfo{author}{\bibfnamefont{H.-J.} \bibnamefont{Zhang}},
  \bibinfo{author}{\bibfnamefont{S.-C.} \bibnamefont{Zhang}},
  \bibinfo{author}{\bibfnamefont{X.}~\bibnamefont{Dai}}, \bibnamefont{and}
  \bibinfo{author}{\bibfnamefont{Z.}~\bibnamefont{Fang}},
  \bibinfo{journal}{Science} \textbf{\bibinfo{volume}{329}},
  \bibinfo{pages}{61} (\bibinfo{year}{2010}{\natexlab{b}}),
  \eprint{https://www.science.org/doi/pdf/10.1126/science.1187485},
  \urlprefix\url{https://www.science.org/doi/abs/10.1126/science.1187485}.

\bibitem[{\citenamefont{Kobayashi}(2011)}]{Kobayashi2011}
\bibinfo{author}{\bibfnamefont{K.}~\bibnamefont{Kobayashi}},
  \bibinfo{journal}{Physical Review B} \textbf{\bibinfo{volume}{84}}
  (\bibinfo{year}{2011}).

\bibitem[{\citenamefont{Pertsova and Canali}(2014)}]{Pertsova2014}
\bibinfo{author}{\bibfnamefont{A.}~\bibnamefont{Pertsova}} \bibnamefont{and}
  \bibinfo{author}{\bibfnamefont{C.~M.} \bibnamefont{Canali}},
  \bibinfo{journal}{New Journal of Physics} \textbf{\bibinfo{volume}{16}},
  \bibinfo{pages}{063022} (\bibinfo{year}{2014}),
  \urlprefix\url{https://doi.org/10.1088%2F1367-2630%2F16%2F6%2F063022}.

\bibitem[{\citenamefont{Pertsova et~al.}(2016)\citenamefont{Pertsova, Canali,
  and MacDonald}}]{Pertsova2016}
\bibinfo{author}{\bibfnamefont{A.}~\bibnamefont{Pertsova}},
  \bibinfo{author}{\bibfnamefont{C.~M.} \bibnamefont{Canali}},
  \bibnamefont{and} \bibinfo{author}{\bibfnamefont{A.~H.}
  \bibnamefont{MacDonald}}, \bibinfo{journal}{Phys. Rev. B}
  \textbf{\bibinfo{volume}{94}}, \bibinfo{pages}{121409}
  (\bibinfo{year}{2016}).

\bibitem[{\citenamefont{Harrison}(1999)}]{Walter1999}
\bibinfo{author}{\bibfnamefont{W.~A.} \bibnamefont{Harrison}},
  \emph{\bibinfo{title}{Elementary Electronic Structure}}
  (\bibinfo{publisher}{World Scientific Pub Co Inc}, \bibinfo{year}{1999}),
  ISBN \bibinfo{isbn}{9810238959},
  \urlprefix\url{https://www.xarg.org/ref/a/9810238959/}.

\bibitem[{\citenamefont{Sancho et~al.}(1984)\citenamefont{Sancho, Sancho, and
  Rubio}}]{Sancho1984}
\bibinfo{author}{\bibfnamefont{M.~P.~L.} \bibnamefont{Sancho}},
  \bibinfo{author}{\bibfnamefont{J.~M.~L.} \bibnamefont{Sancho}},
  \bibnamefont{and} \bibinfo{author}{\bibfnamefont{J.}~\bibnamefont{Rubio}},
  \bibinfo{journal}{Jour. of Phys. F: Met. Phys.}
  \textbf{\bibinfo{volume}{14}}, \bibinfo{pages}{1205} (\bibinfo{year}{1984}),
  \urlprefix\url{https://doi.org/10.1088/0305-4608/14/5/016}.

\end{thebibliography}
